\documentclass[preprint, preprintnumbers ,amsmath,amssymb]{revtex4}

\usepackage{graphicx}
\usepackage{dcolumn}
\usepackage{bm}

\begin{document}




\title{ Determination of Wave Function Functionals: The Constrained-Search---Variational Method}


\author{Xiao-Yin Pan}
\author{ Viraht Sahni}
\author{Lou Massa}
\affiliation{ The Graduate School of the City University of New
York, New York, New York 10016. }

\date{\today}

\begin{abstract}
In a recent paper [Phys. Rev. Lett. \textbf{93}, 130401 (2004)],
we proposed the idea of expanding the space of variations in
variational calculations of the energy by considering the
approximate wave function $\psi$ to be a functional of functions $
\chi: \psi = \psi[\chi]$ rather than a function.  The space of
variations is expanded because a search over the functions $\chi$
can in principle lead to the true wave function. As the space of
such variations is large, we proposed the constrained-search---
variational method whereby a constrained search is first performed
over all functions $\chi$ such that the wave function functional
$\psi[\chi]$ satisfies a physical constraint such as normalization
or the Fermi-Coulomb hole sum rule, or leads to the known value of
an observable such as the diamagnetic susceptibility, nuclear
magnetic constant or Fermi contact term. A rigorous upper bound to
the energy is then obtained by application of the variational
principle.  A key attribute of the method is that the wave
function functional is accurate throughout space, in contrast to
the standard variational method for which the wave function is
accurate only in those regions  of space contributing principally
to the energy.  In this paper we generalize the equations of the
method to the determination of arbitrary Hermitian single-particle
operators as applied to two-electron atomic and ionic systems. The
description is general and applicable to both ground and excited
states. A discussion on excited states in conjunction with the
theorem of Theophilou is provided.  Here we construct new
analytical 3-parameter ground state  wave function functionals for
the negative ion of atomic Hydrogen and the Helium atom through
the constraint of normalization. We present the results for the
total energy $E$, the expectations of the Hermitian
single-particle operators   $W = \sum_{i} r_{i}^{n} , n = -2,-1,
1, 2, W = \sum_{i} \delta({\bf r}_{i})$, and $W = \sum_{i}
\delta({\bf r}_{i} - {\bf r})$, the structure of the nonlocal
Coulomb hole charge $\rho_{c}({\bf r} {\bf r}')$ as a function of
electron position ${\bf r}$, and the expectations of the two
particle operators $u^{2}, u, 1/u, 1/u^{2}$, where $u = |{\bf
r}_{i} - {\bf r}_{j}|$. The results for all the expectation values
are remarkably accurate when compared with the $1078$-parameter
wave function of Pekeris, thereby demonstrating the accuracy of
the wave function functionals \emph{throughout} space.  The same
accuracy is exhibited for the Coulomb holes when compared with the
`exact' ones determined by Slamet and Sahni. The point of the
improved spatial accuracy of these functionals is further made by
comparisons with the results of other wave functions that are not
functionals such as those due to Hartree-Fock theory and
Caratzoulas-Knowles.  We conclude by describing our current work
on how the constrained-search---variational method in conjunction
with Quantal density functional theory is being applied to
many-electron atoms.

\end{abstract}


\maketitle

\section{ Introduction}

In recent work \cite{1}, we proposed the idea of expanding the
space of variations in standard variational calculations of the
energy \cite{2}, thereby allowing for an improvement of the energy
in such calculations.
 Equivalently, a required level of accuracy could be achieved with fewer
  variational parameters.  In the traditional application of the variational
   principle, the space of variations is limited  by the choice of analytical
    form for the approximate wave function.  For example,
    if Gaussian or Slater-type orbitals or a linear combination
     of such orbitals is employed in the energy functional, the
     variational space is limited by this choice of functions.
     The proposed manner by which the space of variations can be
     expanded is by considering the wave function $\psi$ to be a functional
     of a set of functions $ \chi: \psi = \psi[\chi]$, rather than a function.
     This permits a greater flexibility for the wave function $\psi[\chi]$ because the
     functions $\chi$ may be chosen such that $\psi[\chi]$ reproduces any well-behaved function.
      In principle, a search over such functions can lead to that function $\chi$ for
       which $\psi[\chi]$ is the true wave function.  \\

The space over which the search for the functions $\chi$ is to be
performed, however, is simply too large for practical purposes,
and a subset of this space must be considered.  We define the
subspace over which the search for the functions $\chi$ is to be
performed  by the requirement that the wave function functional
$\psi[\chi]$ satisfy a constraint.  Typical constraints on the
functional $\psi[\chi]$  are those of normalization, the
satisfaction of the Fermi-Coulomb hole sum rule, the requirement
that it lead to observables such as the electron density, nuclear
magnetic constant, diamagnetic susceptibility, Fermi contact term,
or any other physical property of interest.  With the wave
function functional $\psi[\chi]$ thus determined, a rigorous upper
bound to the energy is obtained by application of the variational
principle. In this way, not only is a particular property of
interest or constarint obtained \emph{exactly}, the energy is also
determined accurately since the variational principle ensures it
is correct to second order in the accuracy of the wave function.
We refer to this method of determining an approximate wave
function as the \emph{constrained-search---variational} method.
The method is general in that it is applicable to both ground and
excited
states.\\

An attribute of constructing a wave function functional
$\psi[\chi]$ via the constrained-search---variational method is
that there is an improvement in the structure of the wave function
throughout all space.  Thus, both single-particle expectations
representative of different parts of space as well as two-particle
expectations involving two different points in space are obtained
accurately.\\

As in standard variational calculations, the satisfaction of constraints
imposed on the wave function functional, while ensuring the exactness of
 a specific property or properties, will nonetheless lead to a less accurate
  upper bound to the energy provided the space of variations remains fixed.
   Any such decrease in the accuracy of the upper bound can, however, be offset
    by an increase in the space of variations.\\

The concept of the wave function $\psi$ as a functional
$\psi[\chi]$  is general in that the space of variations may be
expanded through the functions $\chi$.  The number of functions
$\chi$ are also independent of the electron number $N$.  This
contrasts with the Hartree-Fock theory \cite{3} Slater determinant
$\Phi[\phi_{i}]$ wave function which is also a functional but one
of the $N$-electron spin-orbitals $\phi_{i}$. Furthermore, there
is no variational-flexibility of these spin-orbitals once they
have been determined self-consistently by solution \cite{4} of the
Hartree-Fock equations. The space of variations cannot be expanded
further, and therefore the Hartree-Fock theory wave function
functional cannot be adjusted via the spin-orbitals $\phi_{i}$ to
be the true wave function. Thus, this wave function functional
constitutes a point in the variational space as defined for the
functional $\psi[\chi]$.  The determinantal functional
$\Phi[\phi_{i}]$  is therefore not general in the manner of the
proposed $\psi[\chi]$.\\

In our original work \cite{1} we had noted that the
constrained-search---variational method could be extended to the
determination of arbitrary Hermitian single-particle operators. In
sect.2 we present the equations of this generalization as applied
to the ground state of the negative ion of atomic Hydrogen, the
Helium atom, and its isoelectronic sequence.  The extension of
these ideas to excited states in conjunction with the theorem of
Theophilou \cite{5} is also described.  We had also indicated
various ways by which the results presented in our prior work
could be improved.  One such mechanism was to improve the
prefactor in the correlated-determinantal wave function
functional.  In sect.3 we present the results of the application
of the method with such an improved 3-parameter analytical wave
function functional to the ground state of the negative ion of
atomic Hydrogen and the Helium atom, with normalization as the
constraint.  We present the results for the total energy $E$, the
expectations of the Hermitian single-particle operators  $W =
\sum_{i} r_{i}^{n} , n = -2,-1, 1, 2, W = \sum_{i} \delta({\bf
r}_{i})$, and $W = \sum_{i} \delta({\bf r}_{i} - {\bf r})$, the
structure of the dynamic (nonlocal) Coulomb hole charge
$\rho_{c}({\bf r}{\bf r}')$ as a function of electron position
${\bf r}$, and the expectations of the two particle operators
$u^{2}, u, 1/u, 1/u^{2}$, where $u = |{\bf r}_{i} - {\bf r}_{j}|$.
The results for all the expectation values  are remarkably
accurate when compared with the 1078-parameter wave function of
Pekeris \cite{6}, thereby indicating the accuracy of the wave
function functionals \emph{throughout} space.  The same accuracy
is exhibited in a different way by the comparison of the Coulomb
holes with those of the essentially exact holes determined by
Slamet and Sahni\cite{7}.  The results for the energy and two
particle expectations are far superior to those of Hartree-Fock
theory as expected. However, the single-particle expectations are
essentially equivalent since such expectations within Hartree-Fock
theory are correct to second order \cite{8}. The comparison with
Hartree-Fock theory demonstrates how two square-integrable
normalized antisymmetric wave functions can lead to essentially
 the same electron density\cite{9}, but that one can be significantly superior to the
  other.  Our results are also superior to those of the 3-parameter variational
   Caratzoulas-Knowles wave function \cite{10} that has a similar correlation
    term as ours but is not a functional.  In the concluding section 4, we
    describe our current work on how the ideas of constructing wave function
    functionals are being applied in conjunction with Quantal density
    functional theory\cite{11} to  the many-electron atom.  \\

\section{ Constrained-
search--variational method}

In this section we present the generalization of the
constrained-search---variational method for constraints whereby
typical observables such as the diamagnetic susceptibility,
nuclear magnetic constant, Fermi contact term, and the constraint
of normalization are determined exactly. For the two-electron
systems represented by the negative ion of atomic Hydrogen, the
Helium atom, and its isoelectronic sequence, these properties are
represented by the expectations of the single-particle operators
$W=r_{1}^{2}+r_{2}^{2}$, $W=1/r_{1}+1/r_{2}$, $W=\delta({\bf
r}_{1})+\delta({\bf r_{2}})$, and $W=1$. For these two-electron
systems, the  Hamiltonian in atomic units ($e=\hbar=m=1$)
\begin{equation}
\hat{H}=-\frac{1}{2}\nabla_{1}^{2}-\frac{1}{2}\nabla_{2}^{2}-\frac{Z}{r_{1}}
  -\frac{Z}{r_{2}}+\frac{1}{r_{12}},
\end{equation}
where ${\bf r}_{1}$, ${\bf r}_{2}$ are the coordinates of the two
electrons, $r_{12}$ is the distance between them, and $Z$ is the
atomic number.   We next choose the form of the  wave function
functional to be of the general form
\begin{equation}
\psi[\chi]=\Phi(s,t,u)[1-f(\chi; s,t,u)],
\end{equation}
 with $\Phi(s,t,u)$ a  pre-factor and $f(\chi;
 s,t,u)$ a correlated correction term:
 \begin{equation}
f(s,t,u)=e^{-q u}(1+qu)[1-\chi(q; s,t,u)(1+u/2)],
\end{equation}
where $s=r_{1}+r_{2}, \; t=r_{1}-r_{2}, \;  \; u=r_{12}$, are the
Hylleraas coordinates\cite{12}, and  $q$ is a variational
parameter. Note that \emph{any} two-electron wave function in a
\emph{ground} or \emph{excited} state maybe expressed in this
form. The key to the wave function functional is the determination
of the functions $\chi(q; s,t,u)$. The prefactor
 may be chosen to be of some analytical form with variational
 parameters as in the present work, or the Hartree-Fock theory wave function \cite{4},
 or  determined self-consistently
within the framework of Quantal Density Functional Theory
\cite{11}.\\

For purposes of clarity, and thereby of subsequent analytical ease
of solution, we assume the prefactor to depend only on the
variables $s$ and $t$: $\Phi=\Phi(s,t)$, and for the ground
 $1^{1}S $ state to be of the analytical form \cite{13}
\begin{equation}
 \Phi[\alpha,\beta; s,t]=N e^{-\alpha s}cosh(\beta t)=\frac{N}{2}[e^{-Z_{1} r_{1}}
 e^{-Z_{2}r_{2}}+e^{-Z_{1}r_{2}} e^{-Z_{2}r_{1}}
],
\end{equation}
where different orbitals are allocated to electrons with up and
down spins, $\alpha$ and $\beta$ are variational parameters,
 $Z_{1}=(\alpha-\beta)$, $Z_{2}=(\alpha+\beta)$, and
$N$ is the normalization constant (See the Appendix). (Note that
the normalization of the prefactor is independent of that of the
wave function.)
We further assume that $\chi$ is a function only of the variable
$s$: $\Psi=\Psi[\chi(q;s)]$. (The space of variations could be
expanded further by assuming the function $\chi$ to depend
additionally upon the variable $t$, or still further by a
dependence on $t$ and $u$ as well.)\\

The  wave function functional $\Psi[\chi(q;s)]$ for the ground
state then satisfies
the electron-electron cusp condition which in integral form is \cite{14}, \\
  \begin{equation}
   \Psi ({\bf r}_{1},{\bf r}_{2},...{\bf r}_{N})=
   \Psi ({\bf r}_{2},{\bf r}_{2},{\bf r}_{3},...,{\bf r}_{N})(1+  r_{12}/2 )+
    {\bf r}_{12}\cdot {\bf C}({\bf r}_{2},{\bf r}_{3},...,{\bf r}_{N}),
 \end{equation}
 where ${\bf C}({\bf r}_{2},{\bf r}_{3},...,{\bf r}_{N})$ is an
 unknown vector. The wave function functional also satisfies
  the electron-nucleus cusp condition which is \cite{14},
  \begin{equation}
  \psi({\bf r},{\bf r}_{2}, ...{\bf r}_{N})=\psi(0,{\bf r}_{2}, ...{\bf r}_{N}).(1-Z r)+
   {\bf r}\cdot {\bf a}({\bf r}_{2}, ...{\bf r}_{N}),
  \end{equation}
for $\alpha=2$. Here again
 ${\bf a}({\bf r}_{2}, ...{\bf r}_{N})$ is also an unknown vector.\\

 In terms of the Hylleraas coordinates, the Hermitian single-particle operators noted above and the
normalization operator   may be expressed as $W(s,t)$ where,
respectively, $W(s,t)=(s^{2}+t^{2})/2$, $W(s,t)=\frac{4
s}{s^{2}-t^{2}}$, $W(s,t)=\frac{1}{\pi}[
\frac{\delta(\frac{(s+t)}{2})}{(s+t)^{2}}+\frac{\delta(\frac{(s-t)}{2})}{(s-t)^{2}}]$,
and $W(s,t)=1$. In general, observables can be  represented by
single-particle operators  expressed as $W(s,t)$. The expectation
of the operator $W(s,t)$ which is
\begin{equation}
\langle W \rangle =\frac{\int \Psi^{*}[\chi]W(s,t) \Psi[\chi]
d\tau}{\int \Psi^{*}[\chi] \Psi[\chi] d\tau},
\end{equation}
can on substitution of the wave function functional $\psi[\chi]$
of Eq.(2) be written as
\begin{equation}
 \int |\Phi(\alpha,\beta; s, t)|^{2} [W(s,t)-<W>][f^2(q;s,t,u)-2 f(q;s,t,u)+1]
 d\tau =0.
\end{equation}
Equivalently, Eq.(8) may be rewritten as
\begin{equation}
\int_{0}^{\infty} e^{-2 \alpha s} g(s)ds=0,
\end{equation}
where
\begin{equation}
 g(s)=\int _{0}^{s}du u \int_{0}^{u}dt cosh^{2}(\beta t)[W(s,t)-<W>](s^{2}-t^{2})[f^2(q;s,t,u)-2
 f(q;s,t,u)+1].
\end{equation}
We now assume that the expectation $ \langle W \rangle$ is known
either through experiment or via some accurate calculation
\cite{6}. \\

The next step is the constrained search over functions $\chi(q;s)$
for which the expectation $\langle W \rangle$ of Eq.(7) is
obtained. If the parameter $\alpha$ in Eq.(9) is fixed, then there
exist \textit{many} functions $g(s)$ for which the expectation
$\langle W \rangle$ can be obtained. This corresponds to a large
subspace of wave function functionals (See Ref. 1). On the other
hand, if the parameter $\alpha$ is variable, then the only way in
which Eq.(9) can be satisfied is if
\begin{equation}
g(s)=0,
\end{equation}
 \emph{This is equivalent to
the constrained search of all wave function functionals over the
subspace in which Eq.(9) is satisfied.}\\

 Substitution of $f(\chi; s,t,u)$ into Eq.(11) leads to
 a quadratic equation for the function
$\chi(q;s)$:
\begin{equation}
 a(q,s)\chi(q;s)^{2}+2 b(q,s) \chi(q;s)+c(q,s)=0,
\end{equation}
where

\begin{equation}
 a(q,s)=\int_{0}^{s}du u(1+u/2)^{2}(1+qu)^{2}e^{-2 qu}\int_{0}^{u}dt cosh^{2}(\beta t)(s^{2}-t^{2})[W(s,t)-<W>],
\end{equation}
\begin{equation}
 b(q,s)=-\int_{0}^{s}du(1+u/2)(1+qu)[e^{-2 qu}(1+qu)-e^{-qu}]
\int_{0}^{u}dt cosh^{2}(\beta t)(s^{2}-t^{2})[W(s,t)-<W>],
\end{equation}
\begin{equation}
 c(q,s)=\int_{0}^{s}du [e^{-2 qu}(1+qu)^{2}-2
 e^{-qu} (1+qu)+1] \int_{0}^{u}dt cosh^{2}(\beta t)(s^{2}-t^{2})[W(s,t)-<W>].
\end{equation}\\
Thus, in order to ensure that the wave function functional
$\psi[\chi]$ leads to the exact expectation value $<W(s,t)>$, one
has to solve a quadratic equation for the determination of the
functions $\chi(q; s)$. The subspace thus corresponds to two
points. The two solutions $\chi_{1}(q;s)$ and $\chi_{2}(q;s)$ lead
to two normalized wave functions $\psi[\chi_{1}]$  and
$\psi[\chi_{2}]$ each of which in turn give rise to the exact
expectation $<W(s,t)>$.\\

  For the two normalized wave function functionals as determined above,
  the energy functional  in terms of Hylleraas coordinates which is
\begin{eqnarray}
I[\psi[\chi]] &=& \int\psi^{*}{\hat H}\psi d\tau  \\
&=& 2 \pi^{2}
\int_{0}^{\infty}ds \int_{0}^{s}du \int_{0}^{u} dt \{u
(s^{2}-t^{2}) [ (\frac{\partial \psi}{\partial s})^{2}+
(\frac{\partial \psi}{\partial t})^{2}+(\frac{\partial
\psi}{\partial
u})^{2}]\nonumber \\
& & +2 \frac{\partial \psi }{\partial u}[s (u^{2}-t^{2})
\frac{\partial \psi}{\partial s}+ t (s^{2}-u^{2}) \frac{\partial
\psi}{ \partial
t}] \nonumber \\
&&-[4 Z s u-(s^{2}-t^{2})]\psi^{2}\},
\end{eqnarray}
is then minimized with respect to the parameters $\alpha$, $\beta$
and $q$.
 \\

The above framework presented for the ground  $1^{1}S $ state of
the two electron system is general and also applicable to excited
states. For example,  if one were to consider the excited $2^{3}
S$ triplet state of the Helium atom, one could employ for the
prefactor in Eq.(2) for the wave function functional  $\psi[\chi]$
 the expression $\Phi(\alpha;
s,t)=\sqrt{\frac{2}{3}}(\frac{\alpha^{4}}{\pi})e^{-\alpha s} t$.
Note that in this simplest of choices used for explanatory
purposes, screening effects are ignored. With such a choice, the
procedure to determine the wave function functional $\psi[\chi]$
is the same as described above. In addition, this procedure could
be employed in conjunction with  the theorem of Theophilou
\cite{5} according to which if $\varphi_{1},
\varphi_{2},…,\varphi_{m}$,..., are orthonormal trial functions
for the $m $ lowest eigenstates of the Hamiltonian $H$, having
exact eigenvalues $E_{1}, E_{2}, …E_{m}$,... , then
$\sum_{i=1}^{m} \langle \varphi_{i} |H |\varphi_{i}\rangle \geq
\sum_{i=1}^{m} E_{i}$ . In this way, a rigorous upper bound to the
\emph{sum }of the ground and excited states is achieved. With the
ground state energy known, a rigorous upper bound to the excited
state energy is then determined, while simultaneously a physical
constraint or sum rule is satisfied or an observable
obtained exactly.\\

The description of the constrained-search---variational method
given in this section concerns the determination of wave function
functionals that obtain the expectation value of arbitrary
Hermitian single-particle operators exactly. The functions $\chi$
were assumed to depend only on the Hylleraas coordinate $s$, and
as a consequence, a quadratic equation had to be solved for their
determination. If the variational space is expanded, then one
would have to solve an integral equation for the function $\chi$.
\\

The ideas of the constrained-search---variational method may also
be applied to sum rules involving two-particle properties. For
example, consider the pair-correlation density $g({\bf r} {\bf
r'})$ which is the conditional density at ${\bf r}'$ of all other
electrons, given that one electron is at ${\bf r}$, and which
accounts for  electron correlations due to the Pauli exclusion
principle and Coulomb repulsion. The pair-correlation density for
an N-electron system is defined as
\begin{equation}
g({\bf r} {\bf r'})=\langle \Psi|\sum_{i\neq j} \delta({\bf
r}_{i}-{\bf r}) \delta({\bf r}_{j}-{\bf r})|\Psi\rangle/\rho({\bf
r}),
\end{equation}
and satisfies the sum rule
\begin{equation}
\int g({\bf r} {\bf r'}) d{\bf r'}=N-1,
\end{equation}
for each electron position ${\bf r}$. However, in order to
determine the wave function functional $\psi[\chi]$ that satisfies
this sum rule at each electron position, one must solve an
integral equation for $\chi$. The details of the calculation of
such a wave function functional are to be presented
elsewhere\cite{15}.\\

\section{  Application to the ground state of the Helium atom and the negative
ion of atomic Hydrogen}

\begin{figure}
 \begin{center}
 \includegraphics[bb=6 58 538 750, angle=0, scale=0.7]{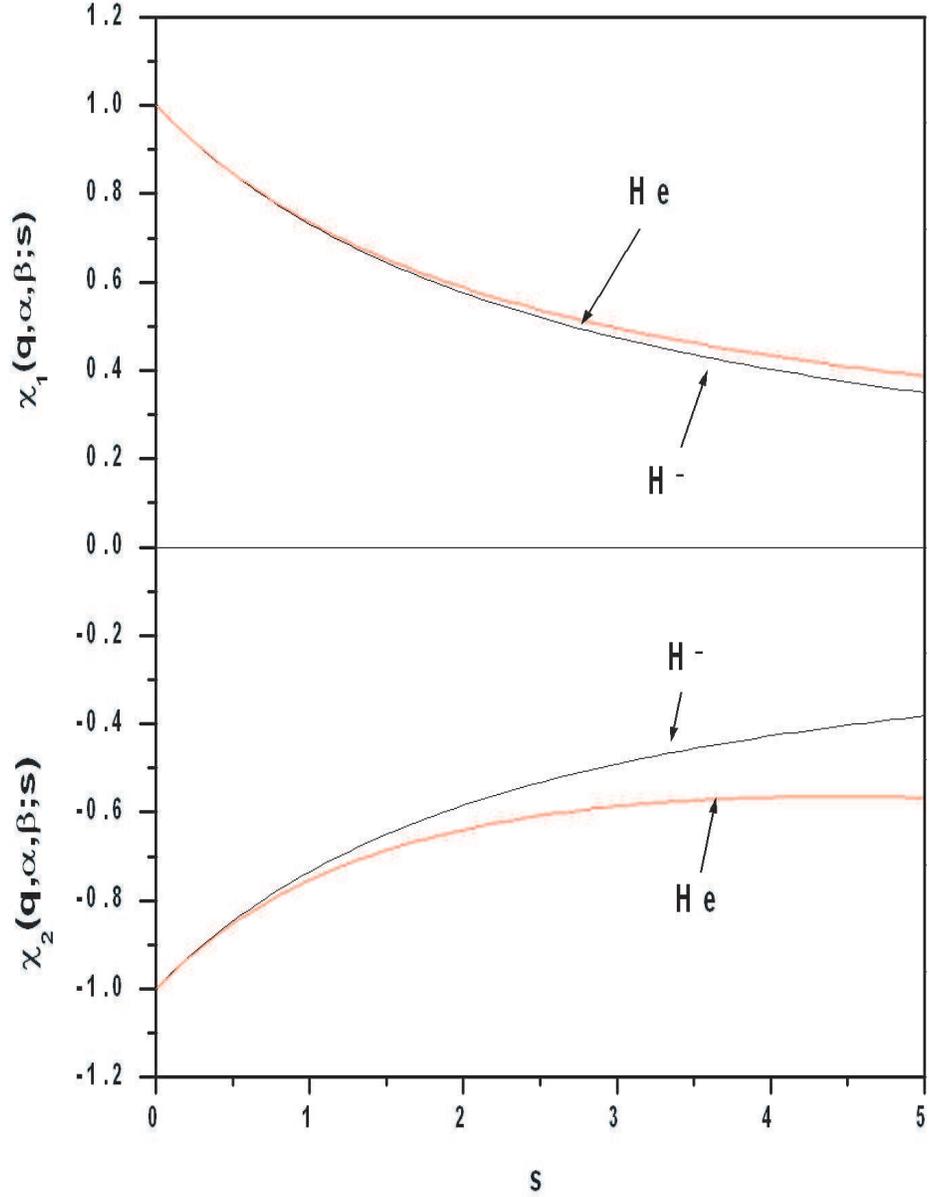}
 \caption{The functions $\chi_{1}(q,\alpha, \beta; s)$ and $\chi_{2}(q,\alpha, \beta; s)$ for
$H^{-}$ and $He$. \label{}}
 \end{center}
 \end{figure}

In this section we apply the constrained-search---variational
method as described above to the ground state of the Helium atom
and the negative ion of atomic Hydrogen.  The constraint employed
is that of normalization, and the prefactor is that of Eq. (4). We
begin with a discussion of the wave function functionals
determined.\\

\emph{Wave function functionals}\\

The 3-parameter wave function functionals are determined by
solution of the quadratic equation Eq.(12).  This solution for the
functions $\chi$ is analytical so that the wave function
functionals $\psi[\chi_{1}]$ and $\psi[\chi_{2}]$ too are
analytical.  We do not provide here the analytical expressions for
$\chi_{1}(q,\alpha,\beta;s)$ and $\chi_{2}(q,\alpha,\beta;s)$, but
these functions are plotted in Fig.1.  Observe that the two
solutions for both $He$ and $H^{-}$ are distinctly different: one
is positive and monotonically decreasing while the other is
negative and monotonically increasing.  Thus, although the two
wave functions have the same structural form, and both satisfy the
normalization constraint and the electron-electron cusp condition,
they are very different.\\

The results as determined by these two wave functions for the
ground state energy, and various single- and two-particle
expectations are given in the subsections below.  Comparisons are
made with the results of the prefactor, Hartree-Fock (HF) theory,
the 3-parameter Caratzoulas-Knowles (CK), and 1078-parameter
Pekeris wave functions.\\

\newpage

\emph{ Ground-state energy}\\

In Table I, we quote the values for the ground-state energy for
 $H^{-}$ and $He$.  The corresponding satisfaction of the virial
  theorem, and percent errors when compared to the values of Pekeris
  for $He$ and those of the variational-perturbation results
  of Aashamar \cite{16} for $H^{-}$ are also given.  Observe that
  the energies obtained by each wave function functional for $H^{-}$ and $He$ are
  an order of magnitude superior to that of the prefactor.  For $He$, these results
   are on the average $0.06\%$ from the Pekeris values.
    They are also an order of magnitude superior to both those of HF and CK.
     For $H^{-}$, both wave function functionals lead to results within $0.1\%$ of
     the Aashamar values, and to positive electron affinities as must
      be the case since the ion is stable.  (In the HF approximation, one does not obtain the negative
      ion of atomic Hydrogen to be stable. The exact satisfaction of the
      virial theorem by HF theory, however, is a consequence of self-consistency.)
      The results clearly demonstrate that highly accurate ground state
      energies can be obtained by constructing few-parameter wave functions
      that are functionals.  These energies are far superior to those determined
      by similar wave functions with
       the same number of parameters but ones that are not
       functionals.\\
\begin {table}

\caption{\label{Table 1.} Rigorous upper bounds to the ground
state energy of $H^{-}$ and $He$ in atomic units as obtained from
the wave function functionals determined via the
constrained-search---variational method, together with the values
due to Hartree-Fock (HF) theory \cite{4}, and the
Caratzoulas-Knowles(CK)\cite{10}, Pekeris\cite{6} and
Aashamar\cite{16} wave functions. The satisfaction of the virial
theorem, and the percent errors compared to the values of Pekeris
and Aashamar are also given. } \vspace{10mm}

\renewcommand{\arraystretch}{1.0}
\begin{tabular}{|c |c |c | c|c | c |}
\hline \hline  Ion or Atom & Wave function & Parameters &
Ground state energy & $\%$ error & $-V/T$  \\
\hline

{$H^{-}$} & {$\Phi$} & $\alpha=0.6612$, $\beta=0.37797$
&$-0.51330$ &
  $2.7374$  & { $2.0001$} \\   \cline{2-6}

&{$\psi[\chi_{1}]$}&  $q=0.274$, $\alpha=0.659$, $\beta=0.308$
&{$-0.52721$}  &{$0.1023$}&
         {$1.9979$} \\ \cline{2-6}

&{$\psi[\chi_{2}]$}& $q=0.094$, $\alpha=0.659$, $\beta=0.306$
&{$-0.52712$}  &{$0.1194$}& {$1.9982$}  \\
\cline{2-6}
       & {Aashamar}  &   & {$-0.52775 $}  &   & {$2.0000$}     \\
       \hline \hline

$He$ & {$\Phi$} & $\alpha=1.68585$, $\beta=0.49732$ &
{$-2.87566$} & {$0.9663$} & { $2.0000$} \\
\cline{2-6}

 &{$\psi[\chi_{1}]$}& $q=0.957$, $\alpha=1.662$, $\beta=0.399$  &{$-2.90158$}  &{$0.0736$}&
        {$1.9975$} \\ \cline{2-6}

         &{$\psi[\chi_{2}]$}& $q=0.242$, $\alpha=1.663$, $\beta=0.399$   &{$-2.90232$}  &{$0.0482$}& {$1.9988$}  \\ \cline{2-6}
& {HF}   &  & {$-2.86168 $}  & {$1.448$}  & { $2.0000$}
\\\cline{2-6}
 & {CK}   &  & {$-2.89007 $}  & {$0.470$}  & { $1.9890$}     \\\cline{2-6}
       & {Pekeris}   &  &{$-2.90372 $}  &    & \ { $2.0000$}     \\
       \hline \hline

\end{tabular}
\end{table}

\emph{Single-particle expectations}\\

In this subsection we present the results of the expectations of
the Hermitian single-particle operators $W =\sum_{i} r_{i}^{n} , n
= -2,-1, 1, 2, W = \sum_{i}\delta({\bf r}_{i})$, and $W =
\sum_{i}\delta({\bf r}_{i}-{\bf r})$. We begin with the
determination of the electron density $\rho({\bf r})$, which is
the expectation of the operator $W = \sum_{i}\delta({\bf
r}_{i}-{\bf r})$, and from which all the other single-particle
expectations may be obtained. (Of course, these expectations may
also be determined directly from the wave function functionals.)
The density $\rho({\bf r})$ is also required for the determination
of the nonlocal Coulomb hole charge distribution $\rho_{c}({\bf
r}{\bf r}')$ as explained in the following subsection.\\

Now the wave function functionals are in terms of the Hylleraas
coordinates $(s,t,u)$ which involve the position of both the
electrons or both their radial distances from the nucleus.  The
electron density $\rho({\bf r})$, on the other hand, depends only
on the coordinates of one of the particles.  Its determination
from wave functions that are written in terms of the Hylleraas
coordinates is as follows.  The electron density
 \begin{equation}
 \rho({\bf r})=\int \psi^{*}(\sum_{i}\delta({\bf r}_{i}-{\bf
 r}))\psi d\tau=2 \int \psi^{2}({\bf r} \; {\bf r}') d{\bf
 r}' ,
 \end{equation}
Using the symmetry of the
 two electronic system, we have
 \begin{equation}
 \int d{\bf r}'=2 \pi \int_{0}^{\infty}r'^{2} dr' \int_{-1}^{1}
 d cos\theta.
 \end{equation}
With $u=\sqrt{r^{2}+r'^{2}-2 r r' cos\theta}$, then, for fixed $r$
and $r'$, we can rewrite Eq.(21)  as
\begin{equation}
 \int d{\bf r}'=2 \pi \int_{0}^{\infty} \frac{r'}{r} dr'
 \int_{|r-r'|}^{r+r'} u du.
 \end{equation}
On rewriting the wave function in terms of ($r,r', u$), and
substituting Eq.(22) into Eq.(20)  leads to
 \begin{eqnarray}
 \rho({\bf r})
 &=& 2 \int_{0}^{\infty}\frac{r'}{r} dr' \int_{|r-r'|}^{r+r'}  u \psi^{2}({r},{r}',
 u) du   \nonumber \\,
 &=& \rho_{0}({\bf r})+\Delta \rho_{0}({\bf r}),
 \end{eqnarray}
 where  $ \rho_{0}({\bf r})$ is the density due to the prefactor
 (see the Appendix for the analytical expression):
\begin{eqnarray}
\rho_{0}({\bf r})& = &2 N^{2} \int e^{-2 \alpha s} cosh^{2}(\beta
t) d {\bf r}'  ,
\end{eqnarray}
 and
\begin{eqnarray}
\Delta\rho_{0}({\bf r})& = &2 N^{2} \int e^{-2 \alpha s}
cosh^{2}(\beta
t) (f^{2}(x;s,t,u)-2 f(\chi; s,t,u))d {\bf r}' ,\nonumber \\
\end{eqnarray}
is the density due to the  correlation term, which can be
evaluated numerically.\\

The electron density at the nucleus is
 \begin{eqnarray}
 \rho(0)&=&\int \psi^{*}(\sum_{i}\delta({\bf r}_{i})\psi d\tau
 \nonumber \\
 &=&\rho_{0}(0)+\Delta \rho_{0}(0) ,
 \end{eqnarray}
where $\rho_{0}(0)$ is the prefactor contribution(see Appendix):
\begin{equation}
\rho_{0}({\bf r})= 2 N^{2} \int e^{-2 \alpha r} cosh^{2}(\beta r)
d {\bf r},
\end{equation}
and the correlation contribution is
\begin{equation}
\Delta\rho_{0}({\bf r}) = 2 N^{2} \int e^{-2 \alpha r'}
cosh^{2}(\beta r') (f^{2}(x;s,t,u)-2 f(\chi; s,t,u))|_{ r_{1}=r=0,
u=r_{2}=r'  }d {\bf r}' .
\end{equation}
\\

In Table II we quote the expectations of the operators $W =
\sum_{i} r_{i}^{n} , n = -2,-1, 1, 2$, and $W = \sum_{i}
\delta({\bf r}_{i})$,  for the ground state of the He atom as
determined by the functionals $\psi[\chi_{1}]$ and
$\psi[\chi_{2}]$ together with those of Hartree-Fock theory, and
the Caratzoulas-Knowles and Pekeris wave functions. The
corresponding percent errors relative to the values of Pekeris are
given in Table III.  As expected (see Table III), the improvement
over the prefactor values is significant.  The results of the two
wave function functionals and those of Hartree-Fock theory are
essentially equivalent, indicating thereby that the corresponding
densities are also essentially the same. The expectations of
single-particle operators in Hartree-Fock theory are, of course,
known to be correct to second order\cite{8}. Hence, both the wave
function functionals are accurate throughout space including the
deep interior and far exterior of the atom. The comparison with
the Caratzoulas-Knowles values (see Table III) is interesting for
its implications.  The wave function functional values are an
order of magnitude superior.  Of course, one does not expect the
CK results to be accurate because these single-particle
expectations are correct only to first order in the accuracy of
the wave function. Thus, our results once again demonstrate, that
wave function functionals determined by the constrained-search---
variational method are superior to variationally determined wave
functions
that are not functionals.\\

\begin{table}
\caption{\label{ } The expectation value of the operator
$W=\sum_{i=1}^{2}r_{i}^{n}; n=-2,-1,1,2 $ and $W= \sum_{i=1}^{2}
\delta({\bf r}_{i})$ for the $He$ atom employing the wave function
functionals determined by the  constraint-search---variational
method, and by the Hartree-Fock theory(HF)\cite{4},
Caratzoulas-Knowles(CK)\cite{10}, and Pekeris \cite{6}  wave
functions (WF). } \vspace{10mm}
\renewcommand{\arraystretch}{1.0}
\begin{tabular}{|c |c |c|c|c|c|}
\hline \hline WF  & $\langle\delta({\bf r}_{1})+\delta({\bf
r}_{2})\rangle$& $<(1/r_{1}+1/r_{2})>$&
$<(1/r_{1}^{2}+1/r_{2}^{2})>$& $<r_{1}+r_{2}>$&
$<(r_{1}^{2}+r_{2}^{2})>$
\\ \hline
$\Phi$ & $3.5025$&  $3.3717$ & $11.930$ &  $1.8758$&$2.4757$  \\
\hline

 $\psi[\chi_{1}]$& $3.6295$ & $3.3750$& $12.033$& $1.8652$ & $2.4156$\\\hline
 $\psi[\chi_{2}]$&$3.6450$ & $3.3735$& $12.048$  & $1.8639$ & $2.4112$\\\hline \hline
 $HF$ & $3.5964$&  $3.3746$ & $11.991$ & $1.8545$ & $2.3697$ \\ \hline
$CK$ & $3.3245$ & $3.3911$ & $11.714$  & $1.7848$& $2.1292$
\\ \hline
$Pekeris$ & $3.62086$& $3.3766$&$12.035$&$1.8589$&$2.3870$\\
\hline \hline
\end{tabular}
\end{table}

\begin{table}
\caption{\label{ } The percentage errors of the results in Table
II relative to the values of Pekeris.} \vspace{10mm}
\renewcommand{\arraystretch}{1.0}
\begin{tabular}{|c |c |c|c|c|c|}
\hline \hline WF  & $\langle\delta({\bf r}_{1})+\delta({\bf
r}_{2})\rangle$& $<(1/r_{1}+1/r_{2})>$&
$<(1/r_{1}^{2}+1/r_{2}^{2})>$ & $<r_{1}+r_{2}>$&
$<(r_{1}^{2}+r_{2}^{2})>$
\\ \hline
$\Phi$ & $3.270$&  $0.145$ & $0.872$  & $0.909$ & $3.716$\\
\hline

 $\psi[\chi_{1}]$& $0.237$ & $0.047$& $0.017$& $0.339$ & $1.198$\\\hline
 $\psi[\chi_{2}]$&$0.667$ & $0.092$& $0.108$  & $0.269$& $1.014$ \\\hline \hline
 $HF$ & $0.676$&  $0.059$ & $0.366$& $0.237$& $0.725$ \\ \hline
$CK$ & $8.185$ & $0.429$ & $2.667$  & $3.986$& $10.800$
\\ \hline

\hline \hline
\end{tabular}
\end{table}

\emph{ Structure of Coulomb holes}\\

We next consider  the structure of the Coulomb hole charge
distribution $\rho_{c}({\bf r}{\bf r}')$ as a function of the
electron position ${\bf r}$.  The definition of this nonlocal or
dynamic charge whose structure changes with electron position for
nonuniform electron gas systems derives from that of the
pair-correlation density $g({\bf r} {\bf r}')$ of Eq.(18) and from
local effective potential energy theory \cite{11}. The
pair-density may be separated into its local and nonlocal
components as
\begin{equation}
g({\bf r} {\bf r}')=\rho({\bf r}')+\rho_{xc}({\bf r} {\bf r}'),
\end{equation}
where $\rho_{xc}({\bf r} {\bf r}')$ is the Fermi-Coulomb hole
charge.  This dynamic charge distribution is the change in the
pair density relative to the density that occurs as a consequence
of the Pauli exclusion principle and Coulomb repulsion. It follows
from Eq.(19) that its total charge is $-1$.  The definition of the
Coulomb hole $\rho_{xc}({\bf r} {\bf r}')$ derives in turn from
that of the Fermi-Coulomb $\rho_{xc}({\bf r} {\bf r}')$ and Fermi
$\rho_{x}({\bf r} {\bf r}')$ holes, the latter being defined
through local effective potential energy theory.  In this theory,
the interacting system as described by the Schr\"{o}dinger
equation is replaced by one of noninteracting Fermions with the
same density. The corresponding wave function is a Slater
determinant of single-particle spin orbitals, and one can then
write down the resulting pair-correlation density $g_{s}({\bf r}
{\bf r}')$ of the model system as
\begin{equation}
g_{s}({\bf r} {\bf r}')=\rho({\bf r}')+\rho_{x}({\bf r} {\bf r}'),
\end{equation}
where $\rho_{x}({\bf r} {\bf r}')$, the Fermi hole, is the
nonlocal component of this pair density, and is a consequence
solely of the Pauli principle. The total charge of the Fermi hole
is also $-1$.  The Coulomb hole is then defined as the difference
between the Fermi-Coulomb and Coulomb holes:
\begin{equation}
\rho_{c}({\bf r} {\bf r}')=\rho_{xc}({\bf r} {\bf
r}')-\rho_{x}({\bf r} {\bf r}'),
\end{equation}
and is thus representative solely of Coulomb correlations.  The
total charge of the Coulomb hole is $0$.  For two-electron systems
in local effective potential theory \cite{11}, the Fermi hole is
then $\rho_{x}({\bf r} {\bf r}')=-\rho({\bf r}')/2$ independent of
electron position ${\bf r}$.\\

\begin{figure}
 \begin{center}
 \includegraphics[bb=0 0 577 797, angle=0, scale=0.7]{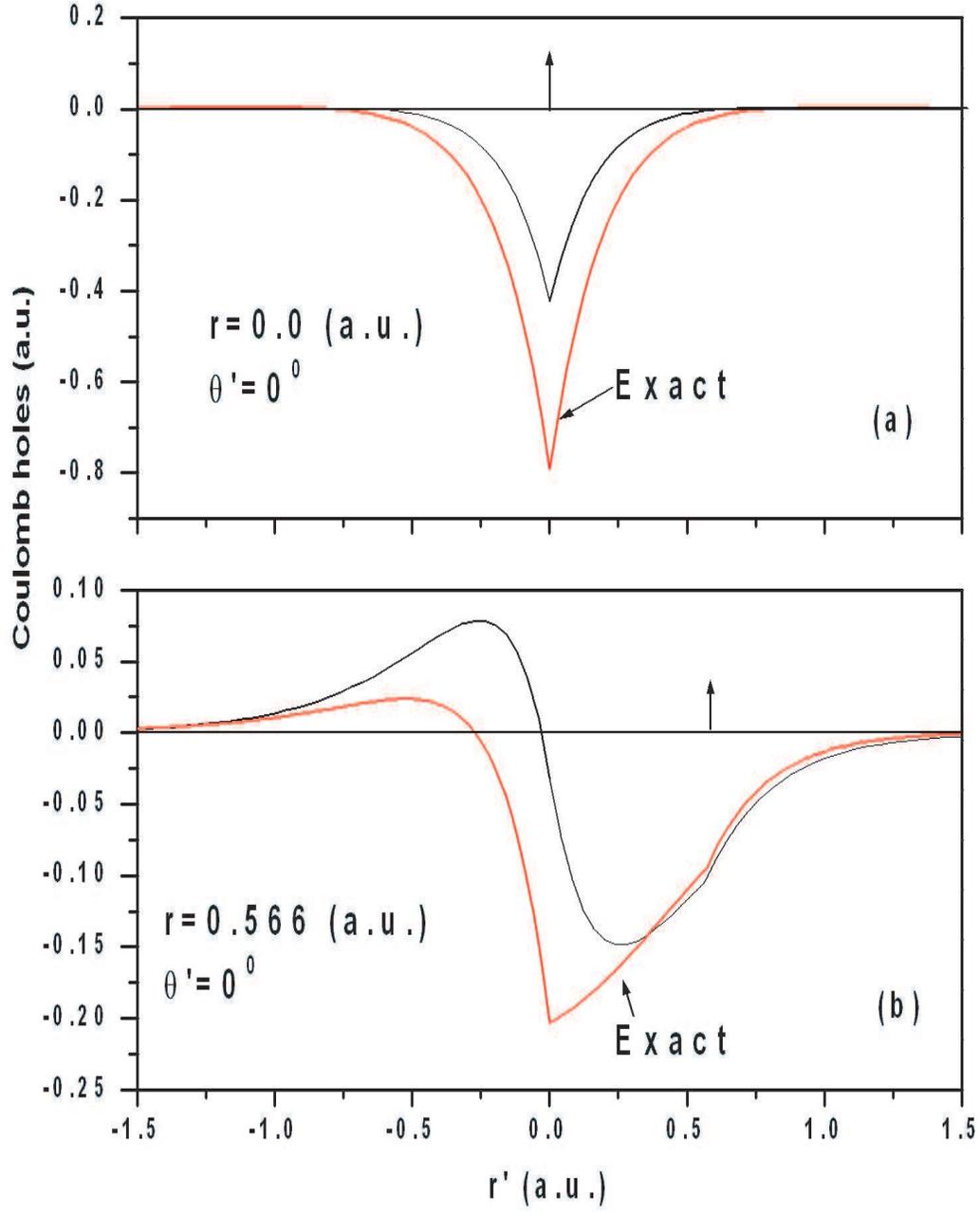}
 \caption{Cross-section through the Coulomb holes determined from the wave function functional $\psi[\chi_{2}]$
 for electron positions at (a)$r=0$ (a.u.), and (b) $r=0.566$ (a.u.).
  The  corresponding  `exact' Coulomb  hole\cite{7} cross sections are also  plotted  for comparison.
  The electron position is indicated by the arrow. \label{}}
 \end{center}
 \end{figure}


\begin{figure}
 \begin{center}
 \includegraphics[bb=0 0 594 788, angle=0, scale=0.7]{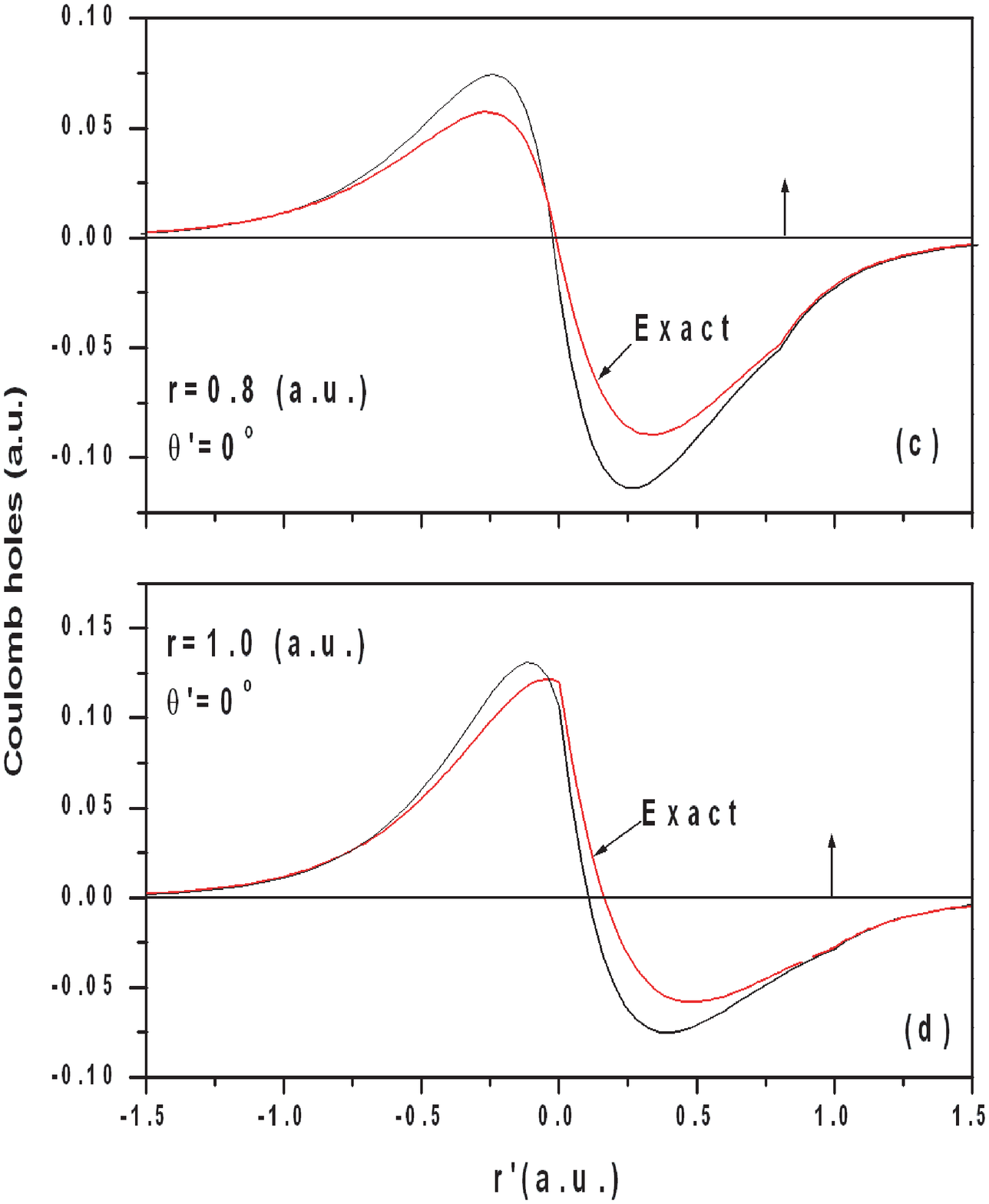}
 \caption{ The figure caption is the same as in Fig.2 except that the cross
sections plotted are for electron positions at (c)$r=0.8$ (a.u.),
and (d) $r=1.0$ (a.u.).\label{}}
 \end{center}
 \end{figure}

\begin{figure}
 \begin{center}
 \includegraphics[bb=0 0 594 788, angle=0, scale=0.7]{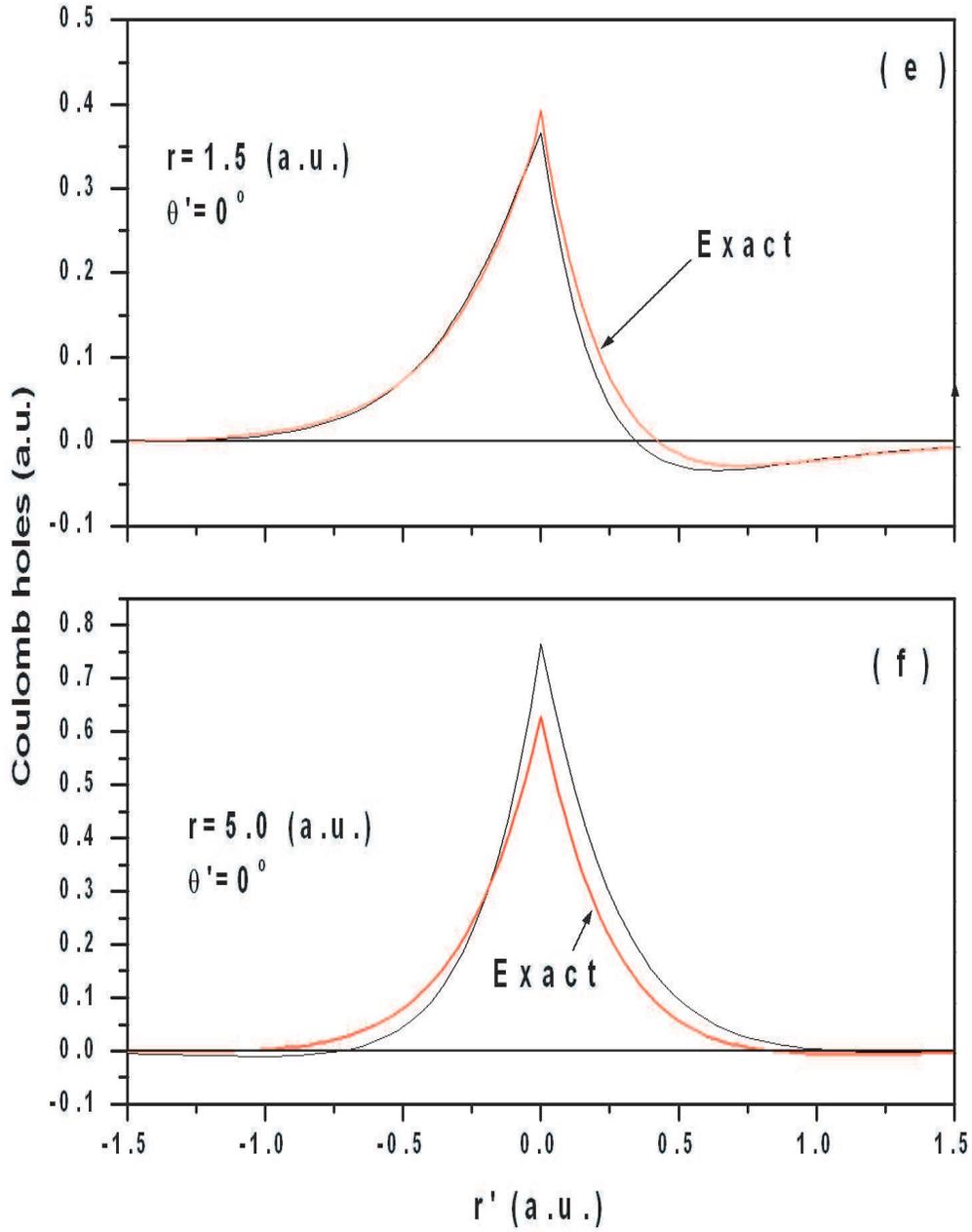}
 \caption{ The figure caption is the same as in Fig.2 except that the cross
sections plotted are for electron positions at (c)$r=1.5$ (a.u.),
and (d) $r=5.0$ (a.u.).\label{}}
 \end{center}
 \end{figure}

In Figs. 2-4, we plot cross sections of the Coulomb hole
$\rho_{c}({\bf r} {\bf r}')$ for different electron positions
${\bf r}$ as obtained via the functional $\psi[\chi_{2}]$ together
with the `exact' Coulomb hole determined by Slamet and
Sahni\cite{7}. (The electron, indicated by the arrow, is on the z
axis corresponding to $\theta = 0^{0}$. The cross section through
the Coulomb hole plotted corresponds to $\theta' = 0^{0}$ with
respect to the electron-nucleus direction. The graph for $r' < 0$
corresponds to the structure for $\theta' = \pi$ and $r' > 0$.)
The electron positions are at $ r = 0, 0.566, 0.8, 1.0, 1.5$, and
$5.0$ (a.u.). It is evident from these figures that the Coulomb
holes as determined from the functional $\psi[\chi_{2}]$ closely
approximate the exact results for electron positions throughout
space: in the interior, within the atom, near its surface and
outside the atom, and in the far asymptotic region. Note the cusp
representative of the electron-electron cusp condition at
 the electron position which is indicated by an arrow in the figures.\\

\emph{ Two-particle expectations}\\

As a consequence of the accuracy of the dynamic Coulomb holes
obtained, we expect the results for the expectation of
two-particle operators to also be accurate.  In Table IV we quote
the values for the expectations of the operators $u^{2}, u, 1/u,
1/u^{2}$, where $u = |{\bf r}_{i} - {\bf r}_{j}|$, together with
the Hartree-Fock and Pekeris values.  The corresponding percent
errors compared to those of Pekeris are given in Table V.  Once
again, the results are an order of magnitude superior to those of
the prefactor, and are accurate for both functionals, although
those due to $\psi[\chi_{2}]$ are consistently superior (see Table
V). Of course, as expected, the Hartree-Fock theory results are
not accurate.\\

If one were able to write the expectation of arbitrary operators
${\hat O}$ as functionals of the density: $\langle {\hat
O}\rangle=\langle \psi[\rho] |{\hat O}|\psi[\rho]\rangle =
O[\rho]$, as is possible in principle according to the
Hohenberg-Kohn theorem \cite{17}, then it is in the expectation of
two-particle operators that the small differences between the
Hartree-Fock theory density and
those of the two wave function functionals would be exhibited.\\


\begin{table}
\caption{\label{ } The expectation value of the operators $u^{2}$,
$u$, $1/u$, $1/u^{2}$, where $u=|{\bf r}_{i}-{\bf r}_{j}|$ as
obtained by the constrained-search---variational method together
with those due to Hartree-Fock theory\cite{4} and the
Pekeris\cite{6} wave function. }\vspace{10mm}

\renewcommand{\arraystretch}{1.0}
\begin{tabular}{|c |c |c|c|c|}
\hline \hline Wave function  &$<u^{2}>$& $<u>$& $<1/u>$ &
$<1/u^{2}>$
\\ \hline
HF  & $2.3694$   & $ 1.3621$  & $ 1.0258$ & $1.8421$
\\ \hline
$\Phi$ &  $2.4757$ & $1.3957$ & $1.0000$ & $1.6998$ \\
\hline
 $\psi[\chi_{1}]$& $2.5325$& $1.4271$& $0.9388$& $1.4300$ \\\hline
 $\psi[\chi_{2}]$& $2.5236$& $1.4241$ & $0.9434$ & $1.4532$ \\\hline \hline
$Pekeris$ & $2.5164$&$1.4220$&$0.9458$&$1.4648$ \\
\hline \hline
\end{tabular}
\end{table}

\begin{table}
\caption{\label{ } The  percentage errors of the results of Table
IV  relative to the values of Pekeris\cite{6}. }\vspace{10mm}

\begin{tabular}{|c |c |c|c|c|}
\hline \hline Wave function  &$\langle u^{2} \rangle$& $\langle u
\rangle$& $\langle 1/u \rangle$ & $\langle 1/u^{2}\rangle $
\\ \hline
HF  & $5.845$   & $ 4.217$  & $ 8.458$ & $25.764$
\\ \hline
$\Phi$ &  $1.618$ & $1.853$ & $5.730$ & $16.045$ \\
\hline
 $\psi[\chi_{1}]$& $0.636$& $0.355$& $0.738$& $2.373$ \\\hline
 $\psi[\chi_{2}]$& $0.286$& $0.142$ & $0.254$ & $0.791$ \\\hline \hline
\end{tabular}
\end{table}


 \section{ Concluding remarks}

  The idea of expanding the space of variations in variational calculations by writing
  the wave function as a functional of functions is appealing not only because the
  functionals lead to more accurate upper bounds for the energy with fewer parameters,
  but also because, as demonstrated in this work, they lead to wave functions that are
  accurate over all space.  Thus, both single- and two-particle expectations are also
  determined accurately.  Certainly, one could claim by comparison with the results of
  Hartree-Fock theory, but without rigorous proof, that single-particle expectations
  obtained thereby are correct to second order in the accuracy of the wave function.
    It is also evident that the accuracy of two-particle expectations lies somewhere
     between first and second order.  In contrast, variationally determined wave functions
      that are not functionals are accurate only in those regions of space contributing to the
       energy.  Thus, for such wave functions, it is the expectation value of only those single- and two-particle
        operators that appear in the
       Hamiltonian that are reasonably accurate.  All other expectations are correct only to first
       order.\\

The results of the present work could be further improved as
follows: by expanding the space of variations through the function
$\chi$; by employing other more efficacious choices for the
analytical form of the correlation factor and thus of the wave
function functional; and by improving the  prefactor. In our work
so far, we have employed analytical forms for the prefactor. ( The
results of our prefactor for the ground state energy of both
$H^{-}$ and $He$ are superior to those of Hartree-Fock theory, see
Table I.) Of course, one could employ the Hartree-Fock theory
Slater determinant as the prefactor. Or one could employ a
determinantal prefactor based on  the orbitals generated within
the local effective potential framework of Quantal density
functional theory (Q-DFT). In principle, these orbitals generate
the true electron density via a model system of noninteracting
Fermions.  The corresponding local potential within Q-DFT depends
upon the wave functions of the interacting and noninteracting
systems. Therefore, the corresponding orbitals generated are
representative of electron correlations due to the Pauli exclusion
principle, Coulomb repulsion, and the correlation
contributions to the kinetic energy.  \\

Finally, we are presently investigating the use of wave function
functionals in conjunction with Q-DFT for the many-electron case
of $N > 2$.  In these calculations, the antisymmetric
determinantal correlated wave function functional employed is of
the form
\begin{equation}
 \psi[\chi]= \Phi\{\phi_{i}\} \Pi_{i\neq j} ( 1 - f( \chi; {\bf r}_{i} , {\bf r}_{j} )).
\end{equation}
 Here $\Phi\{\phi_{i}\}$ is a Slater determinant that defines the
state of the system and whose orbitals $\phi_{i}$  are generated
via the differential equation of Q-DFT, $f(\chi; {\bf r}_{i} ,
{\bf r}_{j} )$ is a spinless correlation functional: $f( \chi;
{\bf r}_{i} , {\bf r}_{j} ) =
e^{-\beta^{2}r^{2}}[1-\chi(R)(1+r/2)]$, where $ {\bf r} = {\bf
r}_{i} - {\bf r}_{j} , {\bf R} = {\bf r}_{i} + {\bf r}_{j} , \beta
= q \rho^{1/3}(R)$, $q $ is a variational parameter, and $\chi(R)$
is determined by the constraint of the Coulomb hole sum rule for
each electron position. This wave function functional satisfies
the electron-electron cusp condition.  In this instance an
integral equation is solved \cite{15} to determine the function
$\chi(R)$. Further, the products of the correlation functional are
limited to lowest order since higher order products of these
factors are less significant \cite{18}. The highest occupied
eigenvalue of Q-DFT differential equation corresponds in principle
to the negative of the ionization potential\cite{11}. The region
that contributes principally to this eigenvalue is the asymptotic
classically forbidden region of the atom. In Q-DFT, the asymptotic
structure of the effective potential is due solely to Pauli
correlations, and can be determined exactly. This is because the
contributions to the potential due to Coulomb correlations and
Correlation-Kinetic effects decay more rapidly than
$(-1/r)$\cite{11}, so that the potential in this region arises
only from the Fermi hole charge which is defined through the
Slater determinant of the orbitals. Thus, accurate ionization
potentials cab be obtained via the use of correlated-determinantal
wave function functionals in conjunction with Q-DFT. These are
variational-self---consistent calculations that lead to upper
bounds for the energy while simultaneously satisfying a nonlocal
physical constraint.  We are also currently investigating the
construction of wave function functionals of the form employed in
the present work, but with the satisfaction of constraints other
than that of
normalization.\\

\appendix*

\section{ }
 We give the analytical expressions for the normalization constant, the
 energy,
and various single- and two-particle  expectation values as
determined by the prefactor wave function
\begin{equation}
 \Phi=N e^{-\alpha s} cosh(\beta t).
\end{equation}

\emph{ Normalization}
\begin{eqnarray}
\int d\tau \Phi^{2}&=&2\pi^{2} N^{2} \int_{0}^{\infty} ds e^{-2
\alpha s} \int_{0}^{s} dt \; cosh^{2}(\beta t) \int_{t}^{s}
du u (s^{2}-t^{2}) \nonumber\\
&=& N^{2}\pi^{2} (\frac{-2 \alpha^{6}+3 \alpha^{4} \beta^{2}-3
\alpha^{2} \beta^{4}+\beta^{6} }{2 \alpha^{6} (\beta-\alpha)^{3}
(\alpha+\beta)^{3}})=1.
\end{eqnarray}

  \emph{Ground-state energy}
\begin{eqnarray}
E_{0} &=& \int\Phi^{*}{\hat H}\Phi d\tau \nonumber \\
&=& 2 \pi^{2} \int_{0}^{\infty}ds \int_{0}^{s}du \int_{0}^{u} dt
\{u (s^{2}-t^{2}) [ (\frac{\partial \Phi}{\partial s})^{2}+
(\frac{\partial \Phi}{\partial t})^{2}+(\frac{\partial
\Phi}{\partial
u})^{2}]\nonumber \\
& & +2 \frac{\partial \Phi }{\partial u}[s (u^{2}-t^{2})
\frac{\partial \Phi}{\partial s}+ t (s^{2}-u^{2}) \frac{\partial
\Phi}{ \partial
t}]-[4 Z s u-(s^{2}-t^{2})]\Phi^{2}\} \nonumber \\
&=& \alpha^{2}-2Z\alpha+ \frac{\alpha (\beta^{2}-\alpha^{2})(10
\alpha^{4}-11 \alpha^{2} \beta^{2}+5 \beta^{4})}{8(-2 \alpha^{6}+3
\alpha^{4} \beta^{2}-3 \alpha^{2} \beta^{4}+\beta^{6}
)} \nonumber\\
& &-\frac{\beta^{4}(3 \alpha^{4}-3 \alpha^{2} \beta^{2}+\beta^{4})
}{(-2 \alpha^{6}+3 \alpha^{4} \beta^{2}-3 \alpha^{2}
\beta^{4}+\beta^{6})}.
\end{eqnarray}

\emph{Expectation values}
\begin{equation}
\rho_{0}({\bf r})= \langle\delta({\bf r}_{1}-{\bf
 r})+\delta({\bf r}_{2}-{\bf
 r})\rangle= N^{2} \pi e^{-2 \alpha
r_{1}}[\frac{1}{\alpha^{3}}+\frac{1}{2} e^{-2 \beta
r_{1}}(\frac{1}{(\alpha-\beta)^{3}}+ \frac{e^{4 \beta
r_{1}}}{(\alpha+\beta)^{3}})],
\end{equation}

 \begin{equation}
 \rho_{0}(0)=\langle\delta({\bf r}_{1})+\delta({\bf r}_{2}
)\rangle=N^{2} \pi [\frac{1}{\alpha^{3}}+\frac{1}{2}
(\frac{1}{(\alpha-\beta)^{3}}+ \frac{1}{(\alpha+\beta)^{3}})],
 \end{equation}

\begin{eqnarray}
\langle r_{1}+r_{2} \rangle =\int d\tau s \Phi^{2}&=&2\pi^{2}
N^{2} \int_{0}^{\infty} ds s e^{-2 \alpha s} \int_{0}^{s} dt\:
cosh^{2}(\beta t) \int_{t}^{s}
du u (s^{2}-t^{2}) \nonumber \\
&=& N^{2}\pi^{2} (\frac{3(2 \alpha^{8}-4 \alpha^{6} \beta^{2}+6
\alpha^{4} \beta^{4}-4 \alpha^{2} \beta^{6}+\beta^{8}) }{2
\alpha^{7} (\beta-\alpha)^{4} (\alpha+\beta)^{4}}).
\end{eqnarray}

\begin{eqnarray}
\langle \frac{1}{r_{1}}+\frac{1}{r_{2}}\rangle=\int d\tau \frac{4
s}{s^{2}-t^{2}} \Phi^{2}&=&2\pi^{2} N^{2} \int_{0}^{\infty} ds 4s
e^{-2 \alpha s} \int_{0}^{s} dt\; cosh^{2}(\beta t) \int_{t}^{s}
du u  \nonumber \\
&=& 2 \alpha.
\end{eqnarray}

\begin{eqnarray}
\langle\frac{1}{r_{1}^{2}}+\frac{1}{r_{2}^{2}}\rangle=\int d\tau
\frac{8 (s^{2}+t^{2}) }{(s^{2}-t^{2})^{2}} \Phi^{2}&=&2\pi^{2}
N^{2} \int_{0}^{\infty} ds  e^{-2 \alpha
s} \int_{0}^{s} dt \;cosh^{2}(\beta t) 4(s^{2}+t^{2})  \nonumber \\
&=&  N^{2}\pi^{2} \frac{(4 \alpha^{6}-4 \alpha^{4} \beta^{2}+6
\alpha^{2} \beta^{4}-2 \beta^{6})}{ \alpha^{4}(\beta-\alpha)^{3}
(\alpha+\beta)^{3}   }.
\end{eqnarray}

\begin{eqnarray}
\langle r_{1}^{2}+r_{2}^{2}\rangle&=&\int d\tau  \frac{
(s^{2}+t^{2}) }{2} \Phi^{2} =  N^{2} \pi^{2}\int_{0}^{\infty} ds
e^{-2 \alpha
s} \int_{0}^{s} dt \; \frac{cosh^{2}(\beta t) (s^{2}+t^{2}) (s^{2}-t^{2})^{2}}{2}  \nonumber \\
&=&  N^{2} \pi^{2} \frac{3(-2 \alpha^{10}+4 \alpha^{8}
\beta^{2}-10 \alpha^{6} \beta^{4}+10 \alpha^{4} \beta^{6}-5
\alpha^{2} \beta^{8}+\beta^{10} )}{ \alpha^{8}(\beta-\alpha)^{5}
(\alpha+\beta)^{5} }.
\end{eqnarray}

\begin{eqnarray}
\langle r_{12}\rangle=\int d\tau  u \Phi^{2}&=&2  N^{2}\pi^{2}
\int_{0}^{\infty} ds e^{-2 \alpha
s} \int_{0}^{s} dt \; cosh^{2}(\beta t)  (s^{2}-t^{2}) \int^{s}_{t} u^{2} du  \nonumber \\
&=& \frac{(70 \alpha^{8}-126 \alpha^{6} \beta^{2}+209 \alpha^{4}
\beta^{4}-140 \alpha^{2} \beta^{6}+35 \beta^{8} )}{16
\alpha(\beta^{2}-\alpha^{2}) (-2 \alpha^{6}+ 3
\alpha^{4}\beta^{2}-3 \alpha^{2} \beta^{4}+\beta^{6}) }.
\end{eqnarray}

\begin{eqnarray}
\langle r_{12}^{2} \rangle=\int d\tau  u^{2} \Phi^{2}&=&2
N^{2}\pi^{2} \int_{0}^{\infty} ds e^{-2 \alpha
s} \int_{0}^{s} dt \; cosh^{2}(\beta t)  (s^{2}-t^{2}) \int^{s}_{t} u^{3} du \nonumber \\
&=&  \frac{6(-2 \alpha^{10}+4 \alpha^{8} \beta^{2}-10 \alpha^{6}
\beta^{4}+10 \alpha^{4} \beta^{6}-5 \alpha^{2}
\beta^{8}+\beta^{10} )}{ \alpha^{2}(\beta^{2}-\alpha^{2})^{2} (-2
\alpha^{6}+ 3 \alpha^{4}\beta^{2}-3 \alpha^{2}
\beta^{4}+\beta^{6}) }.
\end{eqnarray}

\begin{eqnarray}
\langle \frac{1}{r_{12}}\rangle =\int d\tau  \frac{1}{u}
\Phi^{2}&=&2  N^{2} \pi^{2} \int_{0}^{\infty} ds e^{-2 \alpha
s} \int_{0}^{s} dt \; cosh^{2}(\beta t)  (s^{2}-t^{2}) \int^{s}_{t}  du  \nonumber \\
&=&  \frac{\alpha (\beta^{2}-\alpha^{2})(10 \alpha^{4}-11
\alpha^{2} \beta^{2}+5
 \beta^{4} )}{8(-2
\alpha^{6}+ 3 \alpha^{4}\beta^{2}-3 \alpha^{2}
\beta^{4}+\beta^{6}) }.
\end{eqnarray}

\begin{equation}
\langle\frac{1}{r_{12}^{2}}\rangle=\int d\tau  \frac{1}{u^{2}}
\Phi^{2}=2  N^{2}\pi^{2} \int_{0}^{\infty} ds e^{-2 \alpha s}
\int_{0}^{s} dt cosh^{2}(\beta t)  (s^{2}-t^{2}) \int^{s}_{t}
\frac{1}{u} du
\end{equation}
Eq.(A.13) can be evaluated numerically.\\

\begin{acknowledgments}
This work was supported  by the Research Foundation of
 CUNY. L. M. was supported in part by NSF through CREST, and by
 a ``Research Centers in Minority Institutions'' award, RR-03037,
 from the National Center for Research Resources, National
 Institutes of Health. \\
\end{acknowledgments}





\end{document}